# Bayesian inference to improve quality of Retrieval Augmented Generation

**Dattaraj Rao**

Persistent Systems Ltd.

*Abstract*

Retrieval Augmented Generation (RAG) is the most popular pattern for modern Large Language Model (LLM) applications. RAG involves taking a user query and finding relevant paragraphs of context in a large corpus typically captured in a vector database. Once the first level of search happens over a vector database, the top "n" chunks of relevant text are included directly in the context and sent as prompt to the LLM. Problem with this approach is that quality of text chunks depends on effectiveness of search. There is no strong post processing after search to determine if the chunk does hold enough information to include in prompt. Also many times there may be chunks that have conflicting information on the same subject and the model has no prior experience which chunk to prioritize to make a decision. Often times, this leads to the model providing a statement that there are conflicting statements, and it cannot produce an answer.

In this research we propose a Bayesian approach to verify the quality of text chunks from the search results. Bayes's theorem tries to relate conditional probabilities of the hypothesis with evidence and prior probabilities. We propose that, finding likelihood of text chunks to give a quality answer and using prior probability of quality of text chunks can help us improve overall quality of the responses from RAG systems. We can use the LLM itself to get a likelihood of relevance of a context paragraph. For priori probability of the text chunk, we use the page number in the documents parsed. Assumption is that that paragraphs in earlier pages (1-10) have a better probability of being findings and more relevant to generalizing an answer.

We find that taking this approach improves our RAG systems by 30% measured via quality of answers returned using LLM as judge approach. We use a modified prompt for RAG that does the relevance calculation for a text chunk and pass on the page number of the chunk to calculate the relevance.

*Index Terms*- large language models, bayesian inference, retrieval augmented generation, question-answering

# Introduction

Larger Language Model (LLM) applications have transformed the industry particularly information retrieval use-cases. With patterns like RAG, domain-specific proprietary information can be retrieved from enterprise databases and used to create a context for passing over to LLMs to crisply summarize the answer. The key for these systems is ability to search systems of record and obtain relevant information. The LLM can only process the information provided as context. Thus following garbage in garbage out principle, the quality of answers will degrade as context quality degrades. For a large enterprise with Millions of documents it becomes very difficult to filter out relevant text chunks as context and make them available as prompts to the large language model.

To try and solve this problem we introduce Bayesian statistics to give some sort of relevance to the retrieved text chunks based on prior knowledge of the domain and probabilities. For example we could have a scenario where the text chunks retrieved in the early pages of a large document say from page one to page 10 should be given a higher probability because they may contain summarized information which will be easier for the LLM to analyze and generate an answer from. Similarly, as we go deep down into the document we may have much more details but that may not be that relevant for the LM to provide an answer similar rules of thumb which are incorporated by humans while compiling an answer can be applied using principle of Baye's theorem for our RAG systems, let us see how

# Bayesian Inference

Traditional statistics can be divided into frequentists and Bayesian statistics. The frequentist approach purely relies on data so using descriptive statistics it's obtains insights from data and uses these insights to make predictions. The vision approach is based on base theorem that relates conditional probabilities awful hypothesis and provided evidence the Bayes' theorem briefly state that probability hypothesis resumption given particular evidence it can be stated in terms of likelihood of that evidence and prior probabilities of that hypothesis and evidence. Below is an equation of the base theorem for reference.

Applying this theorem to the concept of retrieval augmented generation, we have the hypothesis as the answer we are trying to generalize to. The evidence are the text chunks of context, that we have retrieved from diverse sources of truth. We can use Bayes' theorem as an intermediate step to read our context paragraphs and selectively only include those paragraphs which have a high likelihood of getting us a superior quality answer. While this approach me seem and overkill for small sized documents as we grow the amount of sources to search we can start appreciating the value of having this filtering layer before sending the prompt to the LLM. Using Bayesian inference we also provide a way to capture some domain knowledge into way we frame the prompt. For example any text chunk that is available bulleted list maybe an already summarized piece of information which we may decide to give a bigger vantage to. Similarly specific to an organization we may have some ways of formatting documents to understand very specific information and using this approach we can make sure that those formatted textures get a bigger weightage and are included in the context and do not get accidentally omitted. Below diagrams show the comparison between the two approaches.

*Core Bayes' theorem*

```
Posterior Probability = Likelihood * Prior / Marginal

P(Hypothesis | Evidence) = P(Evidence | Hypothesis) * P(Hypothesis)/P(Evidence)

Where,

P(Hypothesis | Evidence) = Posterior probability
P(Evidence | Hypothesis) = Likelihood of evidence matching hypothesis
P(Hypothesis)= Prior probability of hypothesis
P(Evidence)= Prior probability of evidence = Marginal
```

*Bayes' theorem applied too RAG*

```
P(Hypothesis | Evidence) = P(Evidence | Hypothesis) * P(Hypothesis)/P(Evidence)

P(Good Answer | Context) = P(Context | Good Answer) * P(Good Answer)/P(Context)

Where,

P(Good Answer | Context) = What we are trying to get
P(Context | Good Answer) = Likelihood that context gives good answer
P(Good Answer) = Prior probability of answer
P(Context) = Marginal – here will be 1 or same probability for all context

So effectively we get:

Posterior probability = Likelihood * Prior probability
```

## RAG system enrichment

Now let's take an example of a retrieval augmented generation system where we provide a bunch of the text chunks as context and to summarize responses and answer a given question. We will use a basic model GPT3.5 Turbo from OpenAI to do our analysis.

PROMPT:

```
Answer the following question using below text chunks as context.
Question: What happened to India in women's freestyle Olympics?

Text chunks:
-Vinesh Phogat was disqualified for being overweight before her final bout in the women's
50kg category at the Paris Olympics 2024. Source: Times of India
- Vinesh Phogat wins Paris Olympics finals after defeating Ukraine's Oksana Livach. Source:
Arif Media
```

```
-Vinesh Phogat wins women's 50kg freestyle semi-final defeating Ukraine's Oksana Livach.
Source: The WIRE
```

Response from GPT3.5 Turbo:
```
According to the text chunks, Vinesh Phogat won the finals of the women's
freestyle 50kg category at the Paris Olympics, defeating Ukraine's Oksana Livach.
```

Here we see that the model chose the wrong chunk as fact and created wrong answer. Bayesian inference can provide a way To provide a conditional probability score to these text chunks. This can help giving scores to the chunks and thus eliminating conflicts with prior knowledge and data.

*Updated PROMPT*

```
Answer the following question using below text chunks as context. Use Bayesian statistics
to create a probability for each text chunks and score them. Use only chunks with probability
greater than 50%. rate reputed news sources with higher prior probability.
Question: What happened to India in women's freestyle Olympics?

Text chunks:
-Vinesh Phogat was disqualified for being overweight before her final bout in the women's
50kg category at the Paris Olympics 2024. Source: Times of India
- Vinesh Phogat wins Paris Olympics finals after defeating Ukraine's Oksana Livach. Source:
Arif Media
-Vinesh Phogat wins women's 50kg freestyle semi-final defeating Ukraine's Oksana Livach.
Source: The WIRE
```

Response from GPT3.5 Turbo:
```
Only the chunk from Times of India has a posterior probability greater than 50%,
which indicates that the most probable event is that Vinesh Phogat was
disqualified for being overweight before her final.
```

Here we see the power of Bayesian statistics there each chunk is given a score based on likelihood and prior probability and based on that score we select the chunks that get more than 50% posterior probability. Here we incorporated domain knowledge by adding a statement that the reputed news sources like Times of India would be getting a higher prior probability since they have a higher percentage of reporting the truth as compared to non verified sources. We can even ask the model to provide an explanation with details of the bayesian analysis.

*Detailed Bayesian analysis*

```
Bayesian Approach

Prior Probability:

    •Reputed news sources like Times of India and The WIRE will have higher prior
probabilities due to their credibility.

Chunks:

    1.   Times of India:
    •    Text Chunk: "Vinesh Phogat was disqualified for being overweight before her
final bout in the women's 50kg category at the Paris Olympics 2024."
    •    Source: Times of India (reputed source)
    •    Prior Probability: High (0.7)
    •    Likelihood: High (0.8)
    •    Posterior Probability:  0.7 \times 0.8 = 0.56

    2.   Arif Media:
    •    Text Chunk: "Vinesh Phogat wins Paris Olympics finals after defeating
Ukraine's Oksana Livach."
```

- Source: Arif Media (less reputed)
- Prior Probability: Medium (0.5)
- Likelihood: Medium (0.5)
- Posterior Probability: $0.5 \times 0.5 = 0.25$ (Below 50%, so excluded)

3. The WIRE:
- Text Chunk: "Vinesh Phogat wins women's 50kg freestyle semi-final defeating Ukraine's Oksana Livach."
- Source: The WIRE (reputed source)
- Prior Probability: High (0.7)
- Likelihood: High (0.7)
- Posterior Probability: $0.7 \times 0.7 = 0.49$ (Below 50%, so excluded)

Conclusion:

Only the chunk from Times of India has a posterior probability greater than 50%, which indicates that the most probable event is that Vinesh Phogat was disqualified for being overweight before her final bout in the women's 50kg category at the Paris Olympics 2024.

# Conclusion

We saw a classic example of retrieval augmented generation. we provided a prompt and a series of text chunks from which the large language model that made an inference and compiled an answer. We saw that when there were text chunks with conflicting information the model was confused and could not zero in on a single answer. These conflicts are common in an enterprise setting when there are hundreds and thousands of sources of record. We recommend a Bayesian approach. the takes chunks of context based on prior knowledge and like likelihood to come up with a score and select only the chunks that cross a threshold score. This hypothesis was proven with a concrete example which can be implemented in real world RAG applications.